\documentclass[twocolumn,showpacs,preprintnumbers,amssymb]{revtex4}

\usepackage{graphicx}
\usepackage{epsfig}
\usepackage{dcolumn}
\usepackage{bm}
\def\gsim{\lower0.5ex\hbox{$\:\buildrel >\over\sim\:$}}
\def\lsim{\lower0.5ex\hbox{$\:\buildrel <\over\sim\:$}}

\begin{document}

\preprint{BNL-HET-05/1}

\title{Neutrino masses, mixing and leptogenesis in a 
two Higgs doublet model ``for the third generation''}

\author{David Atwood$^a$}%
\email{atwood@iastate.edu}
\author{Shaouly Bar-Shalom$^b$}
\email{shaouly@physics.technion.ac.il}
\author{Amarjit Soni$^c$}%
\email{soni@bnl.gov}
\affiliation{$^a$Department of Physics and Astronomy, Iowa State University, 
Ames, IA 50011, USA\\
$^b$Physics Department, Technion-Institute of Technology, Haifa 32000, Israel\\
$^c$Theory Group, Brookhaven National Laboratory, Upton, NY 11973, USA}

\date{\today}

\begin{abstract}
We examine neutrino oscillations in a two Higgs
doublet model (2HDM) in which the second doublet couples only to
the third generation right-handed up-fermions, i.e., 
to $t_R$ and to $N_3$ which is the heaviest right-handed Majorana neutrino.
The inherently large $\tan\beta$ of this model
can naturally account for 
the large top mass and, based on a quark-lepton 
similarity ansatz, when embedded 
into a seesaw mechanism it can also account for the observed 
neutrino masses and mixing angles 
giving a very small $\theta_{13}$: 
$-0.017 \lsim \theta_{13} \lsim 0.021$ at 99\% CL, and a very 
restrictive 
prediction for the atmospheric mixing angle: 
$42.9^0 \lsim \theta_{atm} \lsim 45.2^0$ at 99\% CL.
The large value of $\tan\beta$ also sets the mass scale 
of the heaviest right-handed Majorana neutrino $N_3$ and 
triggers successful leptogenesis through a CP-asymmetry in the 
decays of the $N_1$ (lightest right-handed Majorana) 
which is $\tan^2\beta$ enhanced compared to 
the CP-asymmetry obtained in models for leptogenesis with one Higgs 
doublet or in the MSSM. 
This enhancement allows us to relax the lower bound on $M_{N_1}$ 
and consequently also the lower bound on the reheating temperature of 
the early universe.   
\end{abstract}


\maketitle


A monumental discovery of the past decade is that neutrinos
have mass! This discovery is bound to have a significant impact
on our understanding of the universe. It also presents us with a
pressing challenge: unraveling of the mixing matrix in the lepton
sector. When we recapitulate the analogous quark case we can easily
grasp the importance and the difficulties of this new task. Indeed, an
obvious but nonetheless crucial question is the relation of the mixing
matrix in the lepton sector to that in the quark sector.
Intensive experimental effort is underway and much more is being
planned, round the globe, to address these issues of vital
importance.

In this work we suggest that these impressive findings in the lepton
sector
are closely related to another remarkable finding of the 90's in
the quark sector, namely that the top quark is enormously heavy
compared to all the other quarks. Specifically, we will propose
a simple, grounds-up approach which treats the 3rd generation
neutrino in a completely analogous manner to the top quark.
This reasoning leads us to a concrete framework for neutrino
masses and mixings with considerable predictive power. For example,
we find that the crucial mixing angle $\theta_{13}$, which
is an important target of many neutrino
experiments, is quite constrained,
$-0.017 \lsim \theta_{13} \lsim 0.021$ at 99\% CL,
in our picture.

Needless to say, these observations of neutrino oscillations \cite{mohap},
confirming the existence of massive
neutrinos, implies that the Standard Model (SM)
can no longer be regarded as a minimal theory
that explains all observed phenomena in Particle Physics.
Bearing this in mind, our objective here is to present a simple
extension of the SM, which will serve as a low energy
effective theory which captures the dominant features of
phenomenology and models some underlying complicated dynamics
of electroweak (EW) symmetry breaking at distances much shorter than
the EW scale.

Recall the allowed $3\sigma$ ranges on the atmospheric and solar
neutrino square mass differences 
and mixing angles (including the CHOOZ $\bar\nu_e$ 
disappearance experiment) \cite{0410030}:
\begin{eqnarray}
7.3 \leq \Delta m_{sol}^2 \cdot 10^{5}~{\rm eV}^2 \leq 9.3 &,&
0.28 \leq \tan^2 \theta_{sol} \leq 0.6 ~,
\nonumber \\
1.6 \leq \Delta m_{atm}^2 \cdot 10^{3}~{\rm eV}^2 \leq 3.6 &,&
0.5 \leq \tan^2 \theta_{atm} \leq 2.1 ~,
\nonumber \\
\sin^2 \theta_{chz} &\leq& 0.041 \label{limits}~,
\end{eqnarray}
\noindent where in the normal hierarchical 
scheme, $m_1 << m_2 << m_3$, we have 
$\Delta m_{sol}^2 = \Delta m_{21}^2$, 
$\Delta m_{atm}^2 = \Delta m_{23}^2$ and 
$\theta_{sol}=\theta_{12},~\theta_{atm}=\theta_{23},
~\theta_{chz}=\theta_{13}$.  

In order to explain the above neutrino mass differences without 
introducing extremely small and unnatural neutrino Dirac 
Yukawa couplings,
the general practice is to add superheavy right handed Majorana neutrino 
fields, $N_i$, and rely on the seesaw mechanism  
for generating sub-eV light neutrinos:
\begin{eqnarray}
{\cal L}={\cal L}_{SM} + {\cal L}_Y(N,L,\phi) + 
M_N^{ij} N_i N_j/2 ~,
\end{eqnarray}
\noindent where, in the minimal setup, 
${\cal L}_Y(N,L,\phi)=Y_{ij}^\nu L_i H N_j $, 
with $L_i=(\nu_{jL},\ell_{jL})$ and
$H$ is the only Higgs doublet
that couples to neutrinos. Then, 
the light neutrino mass matrix is given by the see-saw mechanism 
as: 
\begin{eqnarray}
m_\nu = - m_D M_N^{-1} m_D^T \label{seesawmass}~,
\end{eqnarray}
\noindent with $m_D=v Y^\nu$ and $v=<H>$.

In this work we wish to propose a 2HDM, which we will name the 
``2HDM for the 3rd generation'' (3g2HDM).
The model relates the neutrino sector to the up-quark sector,
and by giving the third generation right-handed neutrino similar 
status as the top quark it provides a natural framework for 
accommodating the observed pattern of neutrino masses and mixing angles. 
In addition, our 3g2HDM 
assumes a specific structure in the Yukawa sector which 
should be viewed as an effective low-energy parametrization of the 
underlying short distance theory. This concrete phenomenological 
framework for the Yukawa sector allows us to successfully address   
both the heaviness of the top-quark and 
the apparent hierarchical structure in the neutrino sector.  

In particular, we extend the 
idea of the so called ``2HDM for the top-quark'' (t2HDM) \cite{das}
to the leptonic sector. 
In the t2HDM one assumes that $\phi_2$ [the second 
Higgs doublet with a much larger vacuum expectation value (VEV)] couples 
{\it only} to 
the top-quark, while the other Higgs doublet $\phi_1$ 
(with a much smaller VEV) couples to all the other fermions.
The large mass hierarchy between the top quark
and the other quarks is then viewed as a consequence of
$v_2/v_1\equiv \tan\beta >> O(1)$,
which, therefore, becomes the ``working assumption'' of the t2HDM.

The t2HDM has several other notable features, such as enhanced 
$H^+bc$ and $H^0 cc$ couplings, new flavor changing 
$H^0tc$ and $H^0 tu$ interactions and new CP phases.     
Thus, the t2HDM can
influence CP-violation in $B$ physics \cite{soni}, 
flavor changing $Z$-decays \cite{ours1} and the production 
of a Higgs in association with a c-jet or a b-jet
in hadron colliders \cite{ours2}.

Following the idea of the t2HDM, we propose the following
Yukawa interactions in the 3g2HDM:
\begin{eqnarray}
{\cal L}_Y =
&-& Y^e {\bar L}_L \phi_1 \ell_R - Y^d {\bar Q}_L \phi_1 d_R
\nonumber \\
&-& Y_1^u {\bar Q}_L \tilde\phi_1 u_R - Y_2^u {\bar Q}_L \tilde\phi_2 u_R
\nonumber \\
&-& Y_1^\nu {\bar L}_L \tilde\phi_1 N - Y_2^\nu {\bar L}_L \tilde\phi_2 N 
+h.c. \label{yukawa} ~,
\end{eqnarray}
\noindent where $N$ are right-handed Majorana neutrinos, $Q$ and 
$L$ are the usual quark and lepton doublets and 
\begin{eqnarray}
Y_1^{u,\nu} \equiv \pmatrix{
a^{u,\nu} & b^{u,\nu} & 0  \cr
a^{u,\nu} & b^{u,\nu} & 0 \cr
0 & \delta b^{u,\nu} & 0 } ~,~
Y_2^{u,\nu} \equiv \pmatrix{
0 & 0 & 0  \cr
0 & 0 & c^{u,\nu} \cr
0 & 0 & c^{u,\nu} }
\label{yun}~,
\end{eqnarray}
\noindent such that, in both the quark and leptonic sectors,
$\phi_2$ couples only to the third generation right-handed 
up-fermions. The Yukawa texture in (\ref{yun}) 
yields the following Dirac neutrino mass matrix (dropping 
the superscript $\nu$):
\begin{eqnarray}
m_D \sim \frac{v_1}{\sqrt{2}} \pmatrix{
a & b & 0  \cr
a & b & c t_\beta \cr
0 & \delta b & c t_\beta } \label{md}~,
\end{eqnarray}
\noindent where $t_\beta \equiv \tan\beta=v_2/v_1$. 

Note that we have assumed the texture zeros 
$(Y_1^{u,\nu})_{13}=0$ and $(Y_2^{u,\nu})_{31}=0$, 
which gives a Dirac neutrino mass matrix in which the upper right and lower 
left entries vanish. This is 
required in order to simplify our discussion below and
to avoid fine tuning in the neutrino Yukawa sector 
when confronted with the experimental limits on neutrino masses 
and mixings (in fact, it is sufficient to 
set only $(Y_2^{\nu})_{13} \to 0$
in order to avoid fine tuning).
The Dirac mass texture in (\ref{md}) is similar to the one which 
arises in the minimal approach  
suggested in \cite{frampton,strumia2}, that is,  
from the requirement of the most economic addition to the SM that can 
accommodate both neutrino oscillation data and the observed baryon asymmetry 
in the universe. Clearly, for such texture zeros to be natural, they  
have to be protected by underlying symmetries of the short distance theory.
For example, SO(10) GUT theories dictate a symmetry
between the up-quark and the Dirac neutrino mass matrices which, 
along with the empirically known properties of the up-quark mass matrix, 
yields the texture zeros in (\ref{md}), see e.g., \cite{0204288}.
In fact, as was shown in \cite{0405016}, there {\it always} exists a scalar 
sector such that {\it any} texture zero in the fermion mass 
matrices can be enforced by means 
of an abelian symmetry once the scalar sector is enlarged to the extent 
required.
Another alternative and attractive framework for the zero 
entries in $Y_1$ and $Y_2$ 
was suggested in \cite{strumia2}: 
such texture zeros may easily result from a vanishingly 
small overlap between the wave 
functions of $N$ and $L$ with the corresponding indexes
in extra-dimensional models. 
The above several possible ways to accommodate the 
desired texture zeros [in (\ref{md})]
do not give us any further insight regarding the underlying reason 
for the structure of the Yukawa couplings in our model.
Nonetheless, they do motivate us to study 
the implications and phenomenology 
of the 3g2HDM with the texture zeros assumed in (\ref{yun}).

We note also that, without loss of generality, further simplifications 
were taken in (\ref{yun}) that allow us to give a more transparent 
analytical derivation of the neutrino oscillations pattern in our 
3g2HDM. In particular, we have chosen 
similar Yukawa couplings within each generation:
$(Y_1^{u,\nu})_{i1}(i=1,2) \sim a^{u,\nu}$,
$(Y_1^{u,\nu})_{i2}(i=1,2,3) \sim b^{u,\nu}$ 
 and $(Y_2^{u,\nu})_{i3}(i=2,3) \sim c^{u,\nu}$.
In addition, a real parameter of $O(1)$ (the parameter $\delta$) 
was added to the $(Y_1)_{23}$ entry 
in (\ref{yun}). We emphasize that these simplifying assumptions have no
effect on our main results below, aside from simplifying the analytical 
formulae given below.
   
In the basis where $M_N$ is diagonal (one can always choose a basis 
for the fields $N$ such that $M_N$ is diagonal),
$M_N=M \cdot diag(\epsilon_{M1},\epsilon_{M2},\epsilon_{M3})$, we 
then obtain from the seesaw mechanism in (\ref{seesawmass}):
\begin{eqnarray}
m_\nu = 
m_\nu^0 \pmatrix{
\epsilon & \epsilon & \delta \bar\epsilon  \cr
\cdot & \epsilon+\omega & \delta \bar\epsilon+\omega \cr
\cdot & \cdot & \delta^2 \bar\epsilon+\omega } \label{mnu}~,
\end{eqnarray}
\noindent where 
\begin{eqnarray}
\epsilon \equiv \frac{a^2}{\epsilon_{M1}}+
\frac{b^2}{\epsilon_{M2}}~,~\bar \epsilon \equiv \epsilon -
\frac{a^2}{\epsilon_{M1}}~,~
\omega \equiv \frac{c^2 t_\beta^2}{\epsilon_{M3}} \label{omega}~,
\end{eqnarray}
\noindent and 
\begin{eqnarray}
m_\nu^0 \equiv (v_1)^2/2M \label{mn0}~.
\end{eqnarray}
In the limit $\omega >> \epsilon,~\bar\epsilon,~\delta$,
the physical neutrino masses and mixing angles are 
then given by \cite{kingmasses}:
\begin{eqnarray}
&&\tan 2 \theta_{23} \sim
\frac{2 r \omega}{\epsilon(\delta^2 -r)},~
\tan 2 \theta_{12} \sim
2 \frac{g}{f},~
\theta_{13} \sim \frac{\epsilon(\delta +r)}{2^{3/2} r \omega}~,
\nonumber\\
&&m_{1} \sim \epsilon m_\nu^0 \left\{1-g \sin 2\theta_{12} +
f \sin^2\theta_{12} \right\}~,
\nonumber\\
&&m_{2} \sim \epsilon m_\nu^0 \left\{1+g \sin 2\theta_{12} +
f \cos^2\theta_{12} \right\},~
\nonumber\\ 
&& m_3 \sim 2 \omega  m_\nu^0 \label{mixing} ~,
\end{eqnarray}
\noindent where we have defined the parameters:
\begin{eqnarray} 
r \equiv \frac{\epsilon}{\bar\epsilon},~ 
g \equiv \frac{\mid r -\delta \mid}{\sqrt{2}r},~ 
f \equiv \frac{\delta^2-2\delta-r}{2r}~.
\end{eqnarray}
\noindent In the following, we will assume normal 
hierarchy for the light neutrinos:
$m_3 >> m_2 >> m_1$, therefore leading to 
$m_3 \sim \sqrt{\Delta m_{atm}^2}$ and 
$m_2 \sim \sqrt{\Delta m_{sol}^2}$. Thus, 
setting for example $2 m_\nu^0 \sim \sqrt{\Delta m_{sol}^2}$,
it follows from the expression for $m_3$ in (\ref{mixing}) that
$\omega \sim \sqrt{\Delta m_{atm}^2/\Delta m_{sol}^2}$. 

Note that, using the definitions for $\omega$ in (\ref{omega}), 
for $m_\nu^0$ in (\ref{mn0}) and the fact that 
$m_t \sim c t_\beta \times (v_1/\sqrt{2})$, our model
yields the following seesaw triple-relation between the heaviest 
light neutrino, the heaviest right-handed Majorana neutrino and the top quark:
\begin{eqnarray}
m_3 \sim 2 \frac{m_t^2}{M_{N_3}} \label{seesawmt}~,
\end{eqnarray}      
Moreover, from (\ref{mn0}) with $2 m_\nu^0 \sim \sqrt{\Delta m_{sol}^2}$
we obtain the typical mass scale ($M$) of the heavy right handed 
neutrinos: 
\begin{eqnarray}
M \sim \frac{v_1^2}{\sqrt{\Delta m_{sol}^2}} \sim 
2 \frac{m_t^2}{t_\beta^2 \sqrt{\Delta m_{sol}^2}} \sim
10^{13}~{\rm GeV}~\label{mscale},
\end{eqnarray}
where, in the last equality, we have taken $m_t \sim v_2/\sqrt{2}$
and $t_\beta \sim O(10)$, which are the ``working assumption'' values  
within the 3g2HDM.

In the following numerical analysis, we will set 
$\omega \sim \sqrt{\Delta m_{atm}^2/\Delta m_{sol}^2} = 5.18$ and 
$m_\nu^0 \sim \sqrt{\Delta m_{sol}^2}/2 = 4.53 \cdot 10^{-3}$ eV, which 
correspond to the best fitted values \cite{0410030}:
$\Delta m_{atm}^2=2.2 \cdot 10^{-3}$ [eV]$^2$ and 
$\Delta m_{sol}^2=8.2 \cdot 10^{-5}$ [eV]$^2$. We note, though,
that the best fitted set of input parameters (obtained by 
performing a minimum $\chi^2$ fit of  
our model to the experimentally measured values of atmospheric and solar 
neutrino masses and mixing angles \cite{0410030}) 
is:\footnote{A more detailed analysis of the free parameter 
space of our model will be presented in \cite{later}.}   
$\omega \sim 5.34$, $\epsilon \sim 0.57$, $\delta \sim -1.28$ and 
$r \sim 1$. 
    
\begin{figure}[htb]
\epsfig{file=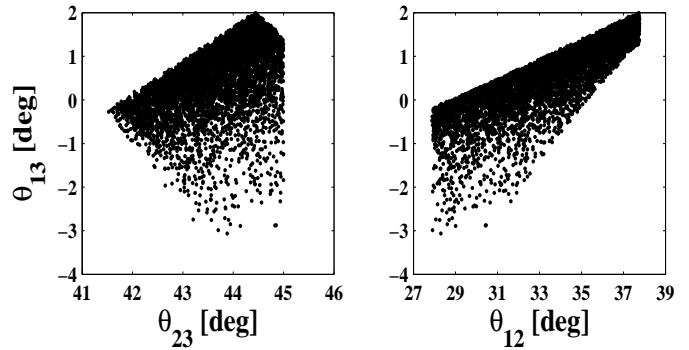,height=5cm,width=10cm}
\vspace{-0.8cm}
\caption{\emph{The allowed ranges in the 
$\theta_{13}-\theta_{23}$ (left) and $\theta_{13}-\theta_{12}$ (right) 
planes, for
$\omega \sim 5.18$ and $m_\nu^0 \sim 4.53 \cdot 10^{-3}$ eV. See also text.}}
\label{fig1}
\end{figure}

In Fig.~\ref{fig1} we give a scatter 
plot of the allowed ranges
in the $\theta_{13}-\theta_{23}$ and $\theta_{13}-\theta_{12}$ planes, i.e., 
subject to the
$3\sigma$ limits listed in (\ref{limits}). This is done
by randomly varying the three input parameters 
$\epsilon,~r$ and $\delta$ 
(requiring $\omega >> \epsilon,~\bar\epsilon,~\delta$) with a 
sample of $3 \cdot 10^6$ points.
We see that the 3g2HDM predicts $\theta_{13}$ to lie 
within (in radians) $-0.05 \lsim \theta_{13} \lsim 0.035$ and 
the atmospheric mixing angle $\theta_{23}$ to be at most a few degrees 
away from maximal. 
In fact, a minimum $\chi^2$ analysis with respect to $\theta_{13}$ and 
$\theta_{23}$ yields: 
\begin{eqnarray}
-0.017 \lsim &\theta_{13}& \lsim 0.021  ~~~~99\% ~{\rm CL} ~, \nonumber \\ 
42.9^0 \lsim &\theta_{23}& \lsim 45.2^0 ~~~~99\% ~{\rm CL} ~,
\end{eqnarray}
with the best fitted value at $\theta_{13} \sim -0.011$ and 
$\theta_{23} \sim 43.9^0$. Notice that this is a rather 
restrictive prediction for the atmospheric mixing 
angle since the experimentally allowed $3\sigma$ range 
for $\theta_{23}$ spans over about 
$20^0$, i.e.,  
$35^0 \lsim \theta_{23}^{exp} \lsim 55^0$ [see (\ref{limits})]. 

\begin{figure}[htb]
\epsfig{file=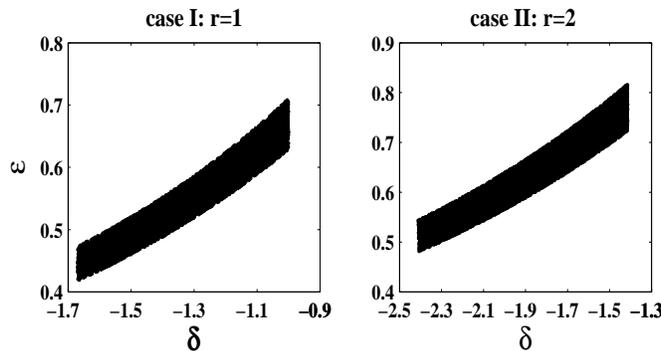,height=5cm,width=10cm}
\vspace{-0.8cm}
\caption{\emph{The allowed ranges in the $\epsilon-\delta$ plane, 
for cases I (left) and II (right). See also caption to Fig.~\ref{fig1}.}}
\label{fig2}
\end{figure}

To better understand what is enforcing maximal 
atmospheric mixing and a very 
small $\theta_{13}$ (as is evident from Fig.~\ref{fig1}), we note that 
from (\ref{mixing}) we have:
\begin{eqnarray} 
\theta_{13} \sim \frac{\epsilon(\delta +r)}{\sqrt{2}} 
\times \tilde\omega^{-1} ~,~ 
\theta_{23} \sim \frac{1}{2} 
\tan^{-1}\left[\frac{\tilde\omega}{\epsilon(\delta^2 -r)} \right] 
\label{relation} ~,
\end{eqnarray}
where 
\begin{eqnarray}
\tilde\omega \equiv 2r\omega \sim 2 r 
\sqrt{\frac{\Delta m_{atm}^2}{\Delta m_{sol}^2}} >> 1 ~,
\end{eqnarray} 
since by definition $r \geq 1$. Therefore, for values of 
$\epsilon$, $\delta$ and $r$ of ${\cal O}(1)$, we have:
\begin{eqnarray}
\mid \theta_{13} \mid \sim \frac{1}{2} 
\sqrt{\frac{\Delta m_{sol}^2}{\Delta m_{atm}^2}} ~,
\end{eqnarray}
as is typical to grand unified (or quark-lepton unified) theories 
\cite{0412050}. Furthermore, from (\ref{relation}) we have (taking 
$\delta < -1$, see discussion below and Fig.~\ref{fig2}):
\begin{eqnarray}
\theta_{23} - \frac{\pi}{4} \sim 
- \frac{1}{2} \sqrt{\frac{\Delta m_{sol}^2}{\Delta m_{atm}^2}} 
\sim - \mid \theta_{13} \mid ~,
\end{eqnarray}   
as can be seen in Fig.~\ref{fig1}.

In Fig.~\ref{fig2} we give a scatter plot of the allowed regions
in the $\delta-\epsilon$ plane, 
in two limiting cases \cite{later}:
\begin{description}
\item{case I:} $r \sim 1$ ($\epsilon \sim \bar\epsilon$),
corresponding to $\frac{a^2}{\epsilon_{M1}}<<\frac{b^2}{\epsilon_{M2}}$.
\item{case II:} $r \sim 2$ ($\epsilon \to 2 \bar\epsilon$),
corresponding to $\frac{a^2}{\epsilon_{M1}} \sim
\frac{b^2}{\epsilon_{M2}}$. 
\end{description} 
We see that case I (corresponding to the best fitted value 
for $r$ \cite{later}) 
is compatible with neutrino 
oscillation data for $0.4 \lsim \epsilon \lsim 0.7$ with 
$-1.7 \lsim \delta \lsim -1$, while the allowed range for case II is
$0.5 \lsim \epsilon \lsim 0.8$ with
$-2.4 \lsim \delta \lsim -1.4$. As an example,
in table~\ref{tab1} we 
give the neutrino mixing angles and masses in cases I and II, for some 
specific values of the allowed parameter space.
\begin{table}[htb]
\begin{center}
\caption[first entry]{Mixing angles and neutrino masses 
in cases I and II 
with $\epsilon=0.5$ and $\epsilon=0.75$, respectively. For both 
cases $\delta=-1.5$, 
$\omega \sim 5.18$ and 
$m_\nu^0 \sim 4.53 \cdot 10^{-3}$ eV. 
\medskip
\protect\label{tab1}}
\begin{tabular}{c||c|c|c|c|c|c}
&\multicolumn{6}{c}{$\delta=-1.5$, $\omega \sim 5.18$, $m_\nu^0 \sim 4.53 \cdot 10^{-3}$ eV} \\
\hline
case $(\epsilon$)& $\theta_{23}$& $\theta_{12}$ &$\theta_{13}$ & $m_1$ [eV]& $m_2$ [eV]& $m_3$ [eV]\\
\hline
I $(0.5)$   & $43^0$ & $29.5^0$ &$-1.14^0$  & $0$ &  $0.0093$ & $0.047$ \\
II $(0.75)$ & $44.7^0$  & $35.9^0$  &$0.69^0$& $3.5\cdot 10^{-4}$ &  $0.0092$ & $0.048$ \\
\end{tabular} 
\end{center}
\end{table}

Let us now examine the consequences of our model 
on the mass spectrum of the heavy Majorana neutrinos, taking into 
account the above constraints coming from neutrino oscillation. 
For definiteness, we will consider case I (i.e., $r \sim 1$, which 
we find as the best fitted value with respect to oscillation data),    
and based on quark-lepton similarity (perhaps motivated by 
GUT scenarios, see e.g., \cite{0204288}), we will take the following 
Ansatz for the neutrino Dirac Yukawa couplings:
\begin{eqnarray}
a \sim O(10^{-3}),~b \sim O(10^{-1}),~c \sim O(1)~\label{abc}.
\end{eqnarray}
Indeed, this choice follows
from the assumption that the Dirac mass matrices of the neutrinos and 
up-quarks exhibit 
a similar hierarchical structure. 
This Ansatz is, therefore, natural in the 3g2HDM under consideration, since 
this model is constructed in order to 
simultaneously explain the large top mass and the hierarchical structure 
of the neutrino masses, by relating the quark sector to the leptonic sector. 
That is, taking $v_1 \sim O(10)$ and $t_\beta \sim O(10)$, 
this Ansatz follows from the up-quark sector; $a^u v_1 \sim O(m_u)$, 
$b^u v_1 \sim O(m_c)$ and $m_t \sim O(c^u v_1 \tan\beta)$, 
if one assumes the quark-lepton similarity:
$a^u \sim a^\nu \sim a$, $b^u \sim b^\nu \sim b$ and 
$c^u \sim c^\nu \sim c$.

It is then easy to extract the mass spectrum of the heavy
Majorana neutrinos in our Ansatz, 
subject to the constraints given in (\ref{limits}). 
In particular, from (\ref{seesawmt}) we have:
\begin{eqnarray}
M_{N_3} \sim 2m_t^2/\sqrt{\Delta m_{atm}^2} \sim 10^{15}~{\rm GeV} 
\label{mn3}~.
\end{eqnarray}
Furthermore, we find that our Ansatz (\ref{abc}) yields 
the following masses for $N_2$ and $N_1$  
[for $r \sim 1$ and for 
$\epsilon \sim O(1)$ in accord with the constrains coming 
from neutrino oscillations, see Fig.~\ref{fig2}]:
\begin{eqnarray}
M_{N_2} \sim 10^{-2}M~~,~~M_{N_1} >> 10^{-6}M  \label{Mmassesa}~,
\end{eqnarray}
\noindent which, for $M \sim 10^{13}$ GeV [see (\ref{mscale})],   
gives $M_{N_2} \sim 10^{11}$ GeV and $M_{N_1} >> 10^{7}$ GeV.

Let us now examine how the 3g2HDM with our quark-lepton similarity 
Ansatz in (\ref{abc}) and with  
$r \sim 1$ and $\epsilon \sim {\cal O}(1)$
fits into 
the mechanism of leptogenesis \cite{leptogen}. Leptogenesis  
is an additional attractive feature, associated with 
the seesaw mechanism, since it allows the possibility of 
generating the observed baryon excess in the universe from 
a lepton asymmetry, driven by decays of the heavy Majorana neutrinos 
\cite{0310123,hambye,leptogen}.
In particular, a CP-asymmetry, $\epsilon_{N_i}$, in the decay 
$N_i \to \ell \phi_j$ 
can generate the lepton asymmetry (see e.g., \cite{0310123,hambye}):
\begin{eqnarray}
n_L/s=\epsilon_{N_i} Y_{N_i}(T>>M_{N_i}) \eta ~,
\end{eqnarray}
\noindent where $Y_{N_i}=n_{N_i}/s$, $n_{N_i}$ being the number density of     
$N_i$ and $s$ the entropy, with $Y_{N_i}(T>>M_{N_i})=135 \zeta(3)/(4\pi^4g_*)$
and $g_*$ being the effective number of spin-degrees of freedom in 
thermal equilibrium. Also,
$\eta$ is the ``washout'' parameter (efficiency factor) that measures
the amount of deviation from the out-of-equilibrium condition at 
the time of the $N_i$ decay.

The lepton asymmetry $n_L/s$ can then be converted into a baryon 
asymmetry through nonperturbative spheleron processes. The conversion 
factor is \cite{convert}: $n_B/s = - (8N_G+4N_H)/(22N_G+13N_H) \times n_L/s$,
where $N_G$ and $N_H$ are the number of generations and Higgs doublets, 
respectively. Thus, in our case, i.e., $N_G=3$, $N_H=2$ and 
$g_* \gsim 100$, we obtain:
\begin{eqnarray}
n_B/s \sim -1.4\times 10^{-3} \epsilon_{N_i} \eta \label{nb}~.
\end{eqnarray}
\noindent As seen from (\ref{mn3}) and (\ref{Mmassesa}), 
our 3g2HDM can lead 
to a hierarchical mass spectrum for the heavy Majorana neutrinos, 
$M_{N_1}<<M_{N_2}<<M_{N_3}$, within a large portion 
of the allowed parameter space.
In particular, imposing $M_{N_1} << M_{N_2}$, 
from (\ref{Mmassesa}) and with $M \sim 10^{13}$ GeV, 
we get: $M_{N_1} \sim 10^{8}-10^{10}$ GeV.
With such an hierarchical mass spectrum, only the CP-asymmetry produced 
by the decay of $N_1$ survives. 

In our model $N_1$ can only decay to the ``light'' doublet $\phi_1$, 
as can be seen from (\ref{yukawa}) and (\ref{yun}). Therefore, the 
CP-asymmetry $\epsilon_{N_1}$ is similar to the one 
obtained in single-Higgs models in the limit $M_{N_1} << M_{N_2},~M_{N_3}$ 
(see e.g., \cite{hambye}):
\begin{eqnarray}
\epsilon_{N_1}= -\frac{3}{16 \pi} \sum_{i=2,3} 
\frac{{\rm Im}\left[ 
\left( {Y_1^\nu}^\dagger Y_1^\nu \right)_{1i}^2 \right]}
{\left( {Y_1^\nu}^\dagger Y_1^\nu \right)_{11}}
\frac{M_{N_1}}{M_{N_j}}  
~,\label{epsn}
\end{eqnarray}
where $Y_1^\nu$ is the Yukawa coupling of $N_1$ to $\phi_1$ given in 
(\ref{yun}). Note that, since $N_3$ does not couple to $\phi_1$ in 
our model, it does not contribute to the sum in 
(\ref{epsn}) above. Therefore, $\epsilon_{N_1}$ is generated only from 
the $i=2$ term in (\ref{epsn}), giving:
\begin{eqnarray}
\epsilon_{N_1}= - \frac{3 b^2}{8 \pi}
\frac{M_{N_1}}{M_{N_2}} \sin 2(\theta_b-\theta_a) \label{eps0}~,
\end{eqnarray}
\noindent where the CP-phases arise from the possible complex 
entries in $Y_1^\nu$: 
$a=|a|e^{i\theta_a}$ and $b=|b|e^{i\theta_b}$.
Thus, using (\ref{mscale}) and $\epsilon \sim b^2/\epsilon_{M2}$ 
(as implied by a fit of the 3g2HDM to neutrino oscillation data, i.e.,
$r \sim 1$), we obtain:
\begin{eqnarray}
\epsilon_{N_1} \sim 
- \frac{3}{16 \pi} 
\frac{t_\beta^2 \sqrt{{\Delta m}_{sol}^2}}{m_t^2} 
\epsilon M_{N_1} \sin 2(\theta_b-\theta_a) \label{eps}~. 
\end{eqnarray}

\noindent It is interesting to note that the above CP-asymmetry 
is a factor of $ \sim t_\beta^2$ larger than the CP-asymmetry 
obtained in models with one Higgs doublet or for that matter 
in SUSY-like models 
where the condensate of a single Higgs 
is responsible for generating the Dirac neutrino  
mass term. Thus, in our 3g2HDM the enhancement factor is of ${\cal O}(100)$ 
since $t_\beta \sim {\cal O}(10)$. To see that we can rewrite 
$\epsilon_{N_1}$ in (\ref{eps0}) as follows:
\begin{eqnarray}
\frac{\epsilon_{N_1}}{\sin 2(\theta_b-\theta_a)} 
\sim - \epsilon_{N_1}^{max} \times \frac{2 \epsilon}{\omega} t_\beta^2 ~,
\end{eqnarray}

\noindent where in our model $\epsilon \sim 0.5$ and $\omega \sim 5$ are  
fixed by oscillation data and $\epsilon_{N_1}^{max}$ is the 
maximum of the CP-asymmetry in models with one Higgs doublet 
(see e.g., \cite{hambye}):
\begin{eqnarray}
\epsilon_{N_1}^{max} = 
\frac{3}{16 \pi} 
\frac{M_{N_1} m_{atm}}{v^2}~,
\end{eqnarray}

\noindent with the atmospheric neutrino mass $m_{atm}=m_3$ given 
in (\ref{mixing}) 
and $v=\sqrt{v_1^2+v_2^2}=246$ GeV.
As mentioned above, this enhancement is possible due to the presence 
of the second Higgs doublet, since in this case 
the light neutrino mass spectrum 
is determined by $v_1$ (the VEV of the lighter Higgs field) and 
not by the EW-scale VEV $v$ (as in the usual scenario 
with one Higgs doublet), and 
the hierarchy $m_{atm} >> m_{sol}$ is dictated by the large $\tan\beta$.
In this way, the CP-asymmetry no longer suffers from a direct relation 
to the Dirac neutrino mass term and is, therefore, not directly 
constraint by 
oscillation data. This observation was also made by 
Fukuyama and Okada in \cite{lepto2HDM} who have investigated leptogenesis 
within a generic two Higgs doublet model.
As will be shown below, this enhancement of the CP-asymmetry 
allows us to reproduce the observed baryon asymmetry in the universe with 
$M_{N_1}$ as low as $10^9$ GeV, which in turn relaxes the 
lower bound on the reheating temperature in the early universe  
to $T_{RH} \gsim 2 \cdot 10^8$ GeV, \cite{hambye}.
            
The efficiency factor, $\eta$, which measures the amount 
of lepton asymmetry left from the CP-asymmetry $\epsilon_{N_1}$, 
is calculated by solving the appropriate Boltzmann equations.
The key parameter entering the Boltzmann equations is the 
``decay parameter'', defined as \cite{hambye}:   
$K \equiv \Gamma_{N_1}/H(T \sim M_{N_1})$, where 
$\Gamma_{N_1}=({Y_1^\nu}^\dagger Y_1^\nu)_{11} M_{N_1}/ 8 \pi$ 
is the total decay width of $N_1$ and 
$H(T) = \sqrt{4 \pi^3 g_*/45}T^2/M_{Planck}$ 
is the Hubble expansion rate. In particular, 
using (\ref{yun}), the decay parameter $K$ is given by (for $g_* \sim 100$):
\begin{eqnarray}
K \sim 4.8 \times 10^{-3} a^2 \cdot \frac{M_{Planck}}{M_{N_1}}~, 
\end{eqnarray}
\noindent which, for $a \sim 10^{-3}$ 
[i.e., Ansatz (\ref{abc})], gives  
$6 \lsim K \lsim 60$ if e.g., $M_{N_1} \sim 10^9 -10^{10}$ GeV.
In this range of values for the decay parameter (corresponding to the 
``mildly strong wash out'' regime), the relation 
$\eta \sim 0.5/K^{1.2}$ constitutes a good fit
to the numerical solution of the Boltzmann equations, 
see e.g., P. Di Bari in \cite{hambye}. Thus, 
using the above fit for $\eta$ and the CP-asymmetry in (\ref{eps}), the 
baryon asymmetry in (\ref{nb}) becomes:
\begin{eqnarray}
\frac{n_B}{s} \sim 10^{-17} 
\frac{t_\beta^2 \sqrt{{\Delta m}_{sol}^2}}{2 m_t^2} 
\epsilon M_{N_1} \left(\frac{M_{N_1}}{{\rm GeV}}\right)^{1.2} 
\sin 2(\theta_b-\theta_a) \label{basym}~. 
\end{eqnarray}  

This has to be compared with the observed baryon to photon 
number ratio 
$n_B/n_\gamma \sim 6 \times 10^{-10}$ \cite{baryon}, implying 
$n_B/s \sim 8.5 \times 10^{-11}$.
For example, taking $\epsilon \sim 0.5$ (see Fig.~\ref{fig2}) and 
${\Delta m}_{sol}^2 \sim 8.2 \cdot 10^{-5}$ eV$^2$, along with
$t_\beta \sim 10$ and $m_t \sim 170$ GeV, (\ref{basym}) 
reproduces the observed baryon asymmetry for e.g., 
$M_{N_1}\sim 10^{10}$ GeV and $\sin 2(\theta_b-\theta_a) \sim 0.1$, or
for $M_{N_1}\sim 10^{9}$ if CP is maximally violated in the sense 
that $\sin 2(\theta_b-\theta_a) \sim 1$.  

Summarizing, we have presented here a simple extension
to the Standard Model, in a grounds-up approach,
which leads us to a framework for neutrino masses and mixings
with considerable predictive power. This can clearly have
significant impact on on-going as well as planned experiments.
To briefly recapitulate, we have constructed a 
2HDM in which the second doublet, with 
a much larger VEV, couples only to
the third generation right-handed up-fermions (the top-quark and the 
3rd generation right-handed neutrino)
and the other doublet couples to all
other fermions. 
Thus, this model is a possible effective low energy parametrization 
of an underlying short distance theory which envisions a close 
relation between
quark dynamics and neutrino physics.
The key parameter of this 2HDM is $\tan\beta$ 
which is assumed to be of $O(10)$ in order to 
naturally accommodate a large mass for the top quark. 
We have shown that the large value of $\tan\beta$ 
in this model is directly responsible for successfully reproducing 
the observed neutrino oscillation data, predicting a very small 
$\theta_{13}$: $-0.017 \lsim \theta_{13} \lsim 0.021$ at 99\% CL, 
a very restricted allowed range for the atmospheric mixing 
angle: $43^0 \lsim \theta_{23} \lsim 45^0$ at 99\% CL, as well as  
successfully reproducing
the observed baryon asymmetry of the universe through leptogenesis. 
In particular, in our model, the CP-asymmetry in the 
$N_1$ decays that drives leptogenesis 
is larger by a factor of ${\cal O}(\tan^2\beta) \sim {\cal O}(100)$ 
than the CP-asymmetry obtained in models for leptogenesis with one Higgs 
doublet or in the MSSM. This enhancement 
allows to relax the lower bound on $M_{N_1}$ 
and accordingly also the lower bound on the reheating temperature of 
the early universe.

S.B.S thanks the
hospitality of the
theory group in Brookhaven National Laboratory where part of 
this study was performed.
This work was supported in part by US DOE Contract Nos.
DE-FG02-94ER40817 (ISU) and DE-AC02-98CH10886 (BNL).

\end{document}